\documentclass{article}

\usepackage{PRIMEarxiv}

\usepackage[utf8]{inputenc} 
\usepackage[T1]{fontenc}    
\usepackage{hyperref}       
\usepackage{url}            
\usepackage{booktabs}       
\usepackage{amsfonts}       
\usepackage{nicefrac}       
\usepackage{microtype}      
\usepackage{lipsum}
\usepackage{fancyhdr}       
\usepackage{graphicx}       
\graphicspath{{media/}}     
\usepackage{amsmath}
\title{Magnetized Strange Stars and Signals of Gravitational Waves
}

\author{
  S. L\'opez P\'erez \\
  Laboratoire Leprince Ringuet \\
  École Polytechnique \\
  Route de Saclay, 91120, Palaiseau, France\\
  \texttt{lopez@llr.in2p3.fr} \\
   \And
  D. Manreza Paret\\
  Facultad de F\'{\i}sica \\
  Universidad de La Habana \\
  San Lázaro y L, Vedado 10400, La Habana, Cuba\\
  \texttt{dmanreza@fisica.uh.cu} \\
  \AND
    A. P\'erez Mart\'inez \\
    Instituto de Cibern\'{e}tica, Matem\'{a}tica y F\'{\i}sica (ICIMAF)\\
    Calle E esq a 15, Vedado 10400, La Habana, Cuba\\
   \texttt{aurora@icimaf.cu} \\
}

\begin{document}
\maketitle

\begin{abstract}
We study the emission of gravitational waves from spheroidal magnetized strange stars for both an isolated slowly rotating star and a binary system. In the first case, we compute the quadrupole moment and the amplitude of gravitational waves that may be emitted. For the binary system, the tidal deformability is obtained by solving simultaneously the system of spheroidal structure equations and the Love number equation. These results are compared with the data inferred from the GW170817 event which is also used to calculate the mass and tidal deformability of the companion star in the binary system. 

Our model supports binary systems formed by magnetized strange stars describing reasonable signals of gravitational waves contrasted with other models of binary systems composed of magnetized hadronic stars and non-magnetized quark stars.
\end{abstract}


\section{Introduction}
Currently, one of the most important and exciting lines of research in astrophysics is related to gravitational waves (GWs). They were directly detected for the first time in 2016 by the Ligo observatory, a signal resulting from the merger of two black holes (GW150914) \cite{Abbott2016}. Similar events followed (GW151226 and GW170104)\cite{Abbott2016b, Abbott2017}. In October 2017, the Ligo and Virgo collaborations announced the amazing first observation of GWs due to the merger of binary Neutron Stars (NSs). Electromagnetic emission from the resulting collision was also observed in multiple wavelength bands. This occurred on August 17, 2017, so the event is known as GW170817 and represents the first time that a cosmic event is observed with both GWs and light \cite{Abbott2017f}. Thus beginning the era of multi-messenger astronomy: the possibility of correlating the signal of the gravitational wave with other signals detected from the same source at almost the same time. 
Multi-messenger events enrich the information about the source and may help to complete the understanding of NSs. The detection of gravitational waves brought new insights into the perception of the Universe, now we can not only see it but also hear it.

Neutron stars can emit GWs if they undergo some kind of deformation, either by companion-induced perturbation in a binary system, rotation, or magnetic field that deviates it from perfect sphericity. \cite{Abadie, Andersson1998, Andersson1999, Bildsten}. In the case of rotating NSs, they will emit at twice their spin frequency and the associated quadrupole deformations are called mountains. So far, determining the spin and orbit parameters of these stars with high accuracy is not possible with the current generation of detectors and would entail a large computational cost, which is expected to be achieved shortly \cite{ligo4}.
Magnetized isolated stars also undergo deformations, becoming spheroids with an associated mass quadrupole \cite{prc}, therefore they are another source of GW and the comparison between theoretical models and observations.

In a binary system of neutron stars, each object develops a quadrupole mass momentum due to the perturbation induced by its companion through the tidal field. This deformation can be quantified through the parameter called tidal deformation or tidal polarizability which depends only on the internal structure of the star, therefore, very important information about the Equation of State (EoS) can be obtained from it. That is why the detection of GWs has established a new channel to study EoS at the high densities describing Neutron Stars.

The composition of the interior of neutron stars is still unknown. Many theories have been formulated, one of the most popular and exotic being the presence of quark matter, either in a stable state (strange quark matter) or forming color superconductor phases, being color-flavor locked (CFL) one of the best candidates \cite{Phddaryel, milva}. In this regard, the reliability of such models can be tested by comparing the proposed models with the observational data from the GW170817 event, especially the mass-radius relations.

Therefore, this work aims to study the gravitational signals from magnetized strange stars. Starting first from the isolated slowly rotating magnetized strange stars (SSs) composed of strange quark matter described within the framework of the MIT Bag model; calculating their quadrupole moment, rotational tidal number, and GW amplitude considering the specific period of Vela and Crab pulsars \cite{bonazzola}. On the other hand, we study a binary system also formed by magnetized SSs and calculate the tidal deformability due to the tidal field between the stars. We compare our results with some observational constraints derived from the tidal deformability inferred from the GW170817 event, the maximum-mass constraints from various known pulsars, and mass-radius estimates derived from the Neutron Star Interior Composition Explorer (NICER) data \cite{nicer}. In addition, we have compared our results with two different works \cite{vivian, Flores}. The first one computes the tidal deformability for magnetized hadronic stars in a binary system \cite{Flores}, meanwhile in \cite{vivian} the authors consider a binary system of non-magnetized stars composed of quark-gluon matter. The constraints imposed by GW170817 are also confronted. However, in our model, we take into account the magnetic field effects on compact objects through a system of spheroidal structure equations.

The paper is organized as follows. In section I we summarize the EoS of magnetized strange stars. Section II presents the stable configurations of spheroidal structure equations for strange stars:  mass and radius are obtained as well as mass quadrupole. From these results and considering that the magnetized stars rotate slowly the tidal deformability associated with the magnetic field and the amplitude of GW is illustrated for isolated strange stars considering that have the period of Crab and Vela pulsar.   
In section III we study the binary system composed of two magnetized strange stars. The theoretical tidal deformability of this system is studied and a comparison with the GW 170817 is done. Finally, we present conclusions and a discussion of our results.

\section{Equation of State of Magnetized Strange Stars}
\label{sec1}

We consider SSs composed of strange quark matter (SQM) and electrons under the action of a uniform and constant magnetic field oriented in the $z$ direction, $\mathbf{B}=(0,0, B)$. Assuming also a pure dipolar configuration (uniform inside the star) for the magnetic field as a first approximation to show how the magnetic field modifies the observables of the stars. This allows us to make direct connections between its microscopic and macroscopic effects on the stars and contributes to an easier physical understanding of our results in a way that could be enlightening when
working with more complex magnetic-field configurations. To obtain the EoS we use the phenomenological MIT bag model \cite{Chodos}, where quarks are considered as quasi-free particles confined to a ``bag'' that reproduces asymptotic freedom and confinement through the $\mathrm{B}_{\mathrm{Bag}}$ parameter binding energy which is added to the energy of the quarks and subtracted from their pressure \cite{Chodos}. In this case, we fix the quark masses and charges to $m_u=2.16$~MeV, $m_d=4.67$~MeV, $m_s=93$~MeV \cite{particle}, $m_e=0.51$~MeV,$e_u=\frac{2}{3}e$ and $e_d=e_s=-\frac{1}{3}e$.

For a magnetized gas of quarks and electrons, the pressure and the energy density are obtained from the thermodynamical potential in \cite{Felipe2008} and considering the equations for stellar equilibrium inside the star \cite{prc}. The magnetized Strange Stars equation of state (EoS) is
\begin{subequations}\label{eos}
	\begin{eqnarray}
		E &=& \sum_{f}\left[\Omega_f+\mu_fN_f\right]+\mathrm{B}_{\mathrm{Bag}}+\frac{B^2}{8\pi},\\
		P_{\parallel} &=& -\sum_{f}\Omega_f-\mathrm{B}_{\mathrm{Bag}}-\frac{B^2}{8\pi},\\
		P_{\perp} &=& -\sum_{f}\left[\Omega_f+B\mathcal{M}_f\right]-\mathrm{B}_{\mathrm{Bag}}+\frac{B^2}{8\pi},\label{epper}
	\end{eqnarray}
\end{subequations}
where 
\begin{equation}\label{potter}
\Omega_f(B,\mu,T)=-\frac{e_fd_fB}{2\pi^2\beta}\int_{0}^\infty dp_3\sum_{l=0}^{\infty}g(l)\ln \left(1+e^{-\beta(\varepsilon_{lf}-\mu)}\right)\left(1+e^{-\beta(\varepsilon_{lf}+\mu)}\right),
\end{equation} 
being $\varepsilon_{lf}^2=\sqrt{p_3^2+2|e_fB|l+m_f^2}$ the spectrum of a magnetically charged fermion, $\mathcal{M}_f=-\partial\Omega_f/\partial B$ is the magnetization, the particle density is $\mathcal{N}_f=-\partial\Omega_f/\partial\mu$. We have designated with $l$ the Landau levels, $d_f$ is the flavor degeneracy factor: $d_e=1$, $d_{u,d,s}=3$ and $g_l=2-\delta_{l0}$ takes into account the spin degeneracy of the fermions for $l\neq0$. In addition, $e_f$, $d_f$, and $m_f$ are the charge, chemical potential, and mass of each particle, respectively.

In Eqs. \eqref{eos} we can easily identify three different contributions. The first one is given by the sum of the thermodynamical quantities of the species and corresponds to the statistical contribution of each kind of particle. The perpendicular pressure \eqref{epper} includes a contribution that comes from the particle magnetization: $-B\mathcal{M}_f$ \cite{Chaichian}. The second terms in the EoS, $\pm\mathrm{B}_{\mathrm{Bag}}$, are those ensuring asymptotic freedom and confinement for quarks \cite{Chodos}. Finally, the last terms in Eqs. \eqref{eos}
are the classical or Maxwell magnetic-field pressures and energy density $P_{\perp}^B=E^B=-P_{\parallel}^B=B^2/8\pi$ \cite{Ferrer}. These terms are included since they also participate in the gravitational stability of the star.

\begin{figure}[h]
	\centering
    \begin{tabular}{cc}
    \includegraphics[width=0.5\linewidth]{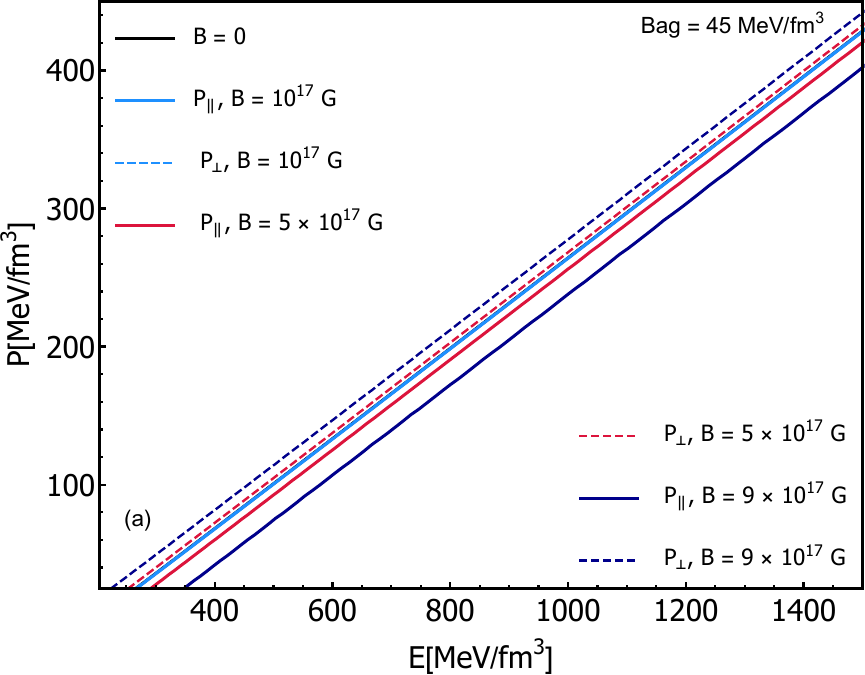}&
    \includegraphics[width=0.5\linewidth]{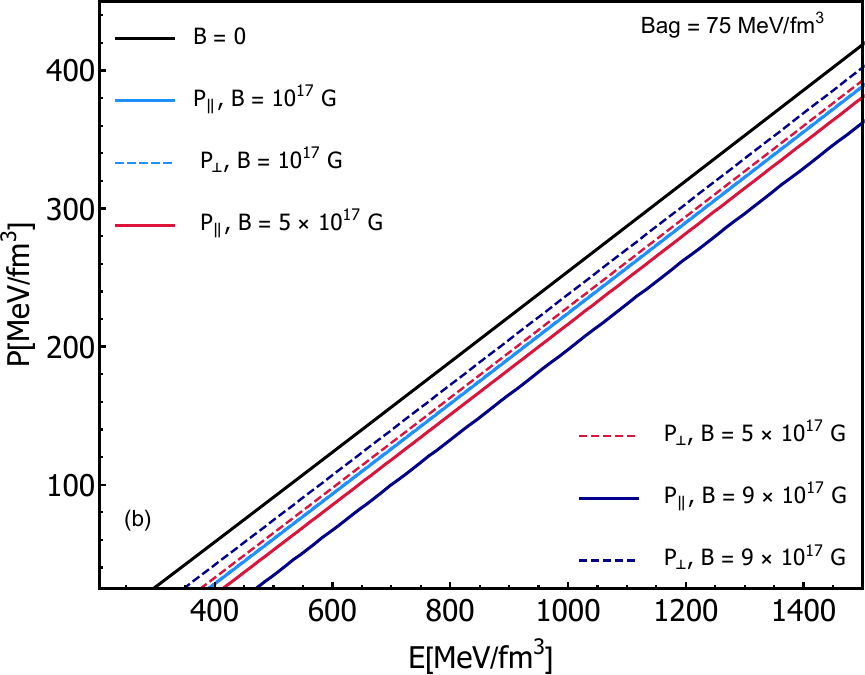}
    \end{tabular}
	\caption{EoS of magnetized SSs at B $=[10^{17}, 5\times10^{17}, 9\times 10^{17}]$~G for $\mathrm{B}_{\mathrm{Bag}}=45$~MeV $/$fm$^3$ and $\mathrm{B}_{\mathrm{Bag}}=75$~MeV $/$fm$^3$.}
	\label{eosplot}
\end{figure}

Figure \ref{eosplot} shows the SSs EoS obtained for magnetic field values of B $=[0,10^{17}, 5\times10^{17}, 9\times 10^{17}]$~G. It is important to note that at higher magnetic field values, the difference between perpendicular and parallel pressures becomes more significant. The $\mathrm{B}_{\mathrm{Bag}}$ parameter affects the EoS, as shown in Figure \ref{eosplot}(a). When $\mathrm{B}_{\mathrm{Bag}}=45$~MeV$/$fm$^3$, the EoS is stiffer than when $\mathrm{B}_{\mathrm{Bag}}=75$~MeV$/$fm$^3$. The effects of magnetic fields and $\mathrm{B}_{\mathrm{Bag}}$ parameters on EoS will be reflected in the macroscopic structure of the star, as we will see in the next section.

\section{Spheroidal structure equations}\label{sec2}

Even when TOV equations, derived from spherical symmetry, are compatible with anisotropic EoS with different tangential and radial pressures\footnote{Actually, isotropic EoS is the simplest assumption to obtain hydrostatic equilibrium equation.}\cite{gleiser} the magnetic anisotropy cannot accommodate to the spherical symmetry. So, it is imperative to derive structure equations within axial symmetry. To include such  effect we use the $\gamma$--structure equations obtained in \cite{prd}
\begin{subequations}\label{gTOV}
    \begin{eqnarray}
    \frac{dM}{dr}&=& 4\pi r^{2}\frac{(E_{\parallel}+E_{\perp})}{2}\gamma, \label{gTOV1}\\
    \frac{dP_{\parallel}}{dz}&=&\frac{1}{\gamma}\frac{dP_{\parallel}}{dr}\nonumber\\
    &=&-\frac{(E_{\parallel}+P_{\parallel})[\frac{r}{2}+4\pi r^{3}P_{\parallel}-\frac{r}{2}(1-\frac{2M}{r})^{\gamma}]}{\gamma r^{2}(1-\frac{2M}{r})^{\gamma}}, \label{gTOV2}\\
    \frac{dP_{\perp}}{dr}&=&-\frac{(E_{\perp}+P_{\perp})[\frac{r}{2}+4\pi r^{3}P_{\perp}-\frac{r}{2}(1-\frac{2M}{r})^{\gamma}]}{r^{2}(1-\frac{2M}{r})^{\gamma}}, \label{gTOV3}
    \end{eqnarray}
\end{subequations}
which describe the variation of the mass and the pressures with the spatial coordinates $r,z$ for an anisotropic axially symmetric compact object as long as the parameter $\gamma = z/r =P_{\parallel 0}/P_{\perp 0}$, where $P_{\parallel 0}$ and $P_{\perp 0}$ are the star central pressures,  is close to one.

Note that Eqs. \eqref{gTOV} are coupled through the dependence with the energy density and the mass. When setting B$=0$, the model automatically yields $P_{\perp}=P_{\parallel}$ and $\gamma=1$. This means that we recover the spherical TOV equations from Eqs. \eqref{gTOV} and thus, the standard non-magnetized solution for the structure of COs. The Eq. \eqref{gTOV} has been used before to describe magnetized Bose-Einstein Condensate (BEC) stars \cite{prc35} and white dwarfs \cite{prd}.

The solutions of the spheroidal structure equation with an equatorial radius $R$ and a polar radius $Z$, i.e. the $M-R$ and $M-Z$ curves correspond to a unique sequence of stable stars, while the $M-R_{\perp}$ and the $M-R_{\parallel}$ ones stand for two different sequences.

\begin{figure}[h]
	\centering
	\begin{tabular}{cc}
		\includegraphics[width=0.5\textwidth]{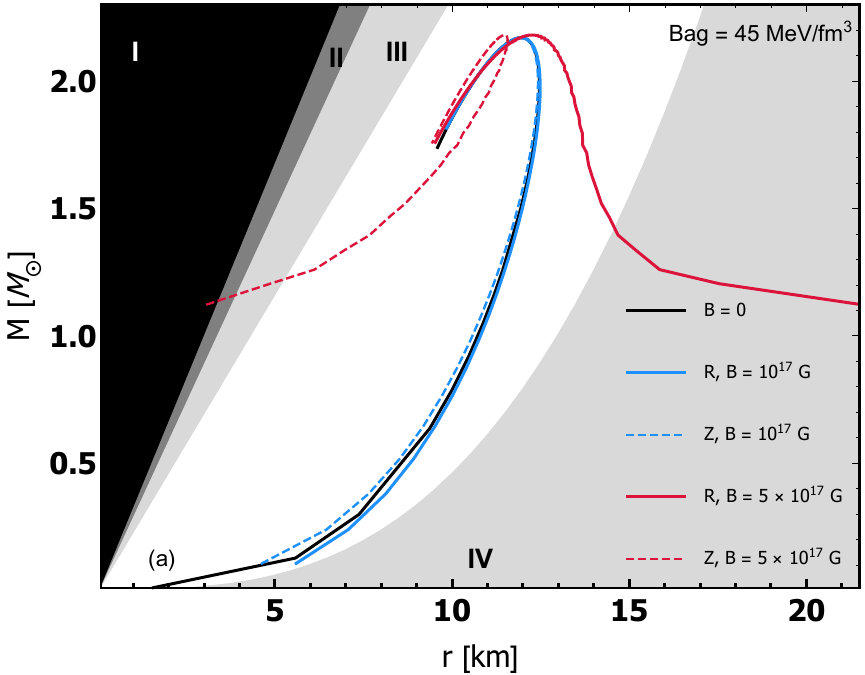}
		& \includegraphics[width=0.5\textwidth]{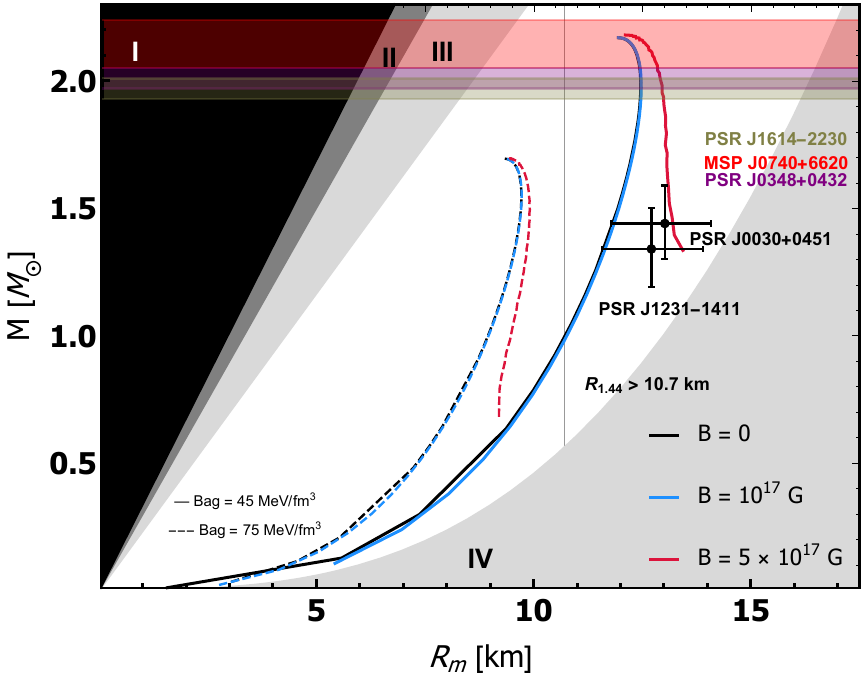}
	\end{tabular}
	\caption{Stable solutions for the spheroidal configurations compared to the non-magnetized case at B $=10^{17}$~G and B $=5\times10^{17}$~G for fixed values $\mathtt{B_{bag}}=45$~MeV$/$fm$^3$ and     $\mathtt{B_{bag}}=75$~MeV$/$fm$^3$ \cite{prc}. Gravitational stability requirement (I). Finite pressure requirement (II). Causality requirement (III). Rotational stability requirement (IV).}
	\label{mrgamma}
\end{figure}

From Fig. \ref{mrgamma} (a), we see that the stars obtained with Eqs. \eqref{gTOV} are oblate objects ($R>Z$), as expected from TOV solutions, where $R_{\perp}>R_{\parallel}$. However, the contrary of what happens with TOV solutions, for which the difference between $R_{\perp}$ and $R_{\parallel}$ increases with the mass, the deformation of $\gamma$--structure equations solutions -the distance between $R$ and $Z$- decreases with the mass.

In Fig. \ref{mrgamma} (b) we show the stable solutions of Eqs. \eqref{gTOV}, taking into account the constraints for the $\gamma$ values analyzed in \cite{prc} and compare with recent observational data for NSs. The low and median bands correspond to objects PSR J$1614-2230$ and PSR J$0348+0432$ with $M = 1.97 \pm 0.04$~M$_{\odot}$ \cite{Demorest} and $M = 2.01 \pm 0.04$~M$_{\odot}$ \cite{Antoniadis}. The upper band represents the result of $2.14^{+0.10}_{-0.09}$~M$_{\odot}$ for the mass of the pulsar MSP J$0740+6620$ at a confidence interval of $68.3\%$ presented in \cite{vivian14}. The range of allowed parameters is further constrained by mass-radius estimates extracted from NICER data, $M=1.44^{+0.15}_{-0.14}$~M$_{\odot}$ with $R=13.02^{+1.24}_{-1.06}$~km \cite{Miller}, $M=1.34^{+0.15}_{-0.16}$~M$_{\odot}$ with $R=12.71^{+1.14}_{-1.19}$~km \cite{Riley} y $R_{1.44}>10.7$~km \cite{Bogdanov}. These estimates are indicated by the black dots with their corresponding error bars. Determination by the NICER group for the neutron star PSR J$0030+0451$ is probably the most reliable measurement at present. As noted, it predicts a radius of about $11$~km-$13$~km for $M\sim 1.4$~M$_{\odot}$. This range for $R$ is on the \textquotedblleft high\textquotedblright side of the expected values. Reports on small radii have been presented over the years (see, for example, references \cite{vivian52, vivian53}), although they involve some modeling and are not as straightforward. The maximum masses of the curves for $\mathtt{B_{bag}}=45$~MeV$/$fm$^3$ lie in the colored bands, while lower masses reach the ranges determined for the other two pulsars.

\subsection{Gravitational Waves for isolated slow rotating magnetized Strange Stars}

General Relativity tells us that only objects with mass quadrupole moment different from zero can emit GWs \cite{Thorne1967}. Mathematically, it is the consequence of matching internal and external Einstein's field equations solutions considering boundary conditions at the surface of the object. Therefore, the amplitude of the GW  is proportional to the variation in time of the object's mass quadrupole \cite{Thorne1967}. Then, GWs are expected from cataclysmic events such as collisions, rotation, or magnetic fields of compact objects that imply deformation far from sphericity. Magnetized stars, described as spheroids, are deformed, having a non-zero quadrupole moment.

In the framework of our structure equations the quadrupole moment of the magnetized SSs is given by \cite{prc42}
\begin{equation}\label{mqf}
Q^{\gamma} = \frac{\gamma}{3}M^3\left(1-\gamma^2\right),
\end{equation}
where $\gamma = 1$ implies $Q=0$, corresponding  to the spherical case.

\begin{figure}[h]
	\centering
	\begin{tabular}{cc}
	\includegraphics[width=0.7\textwidth]{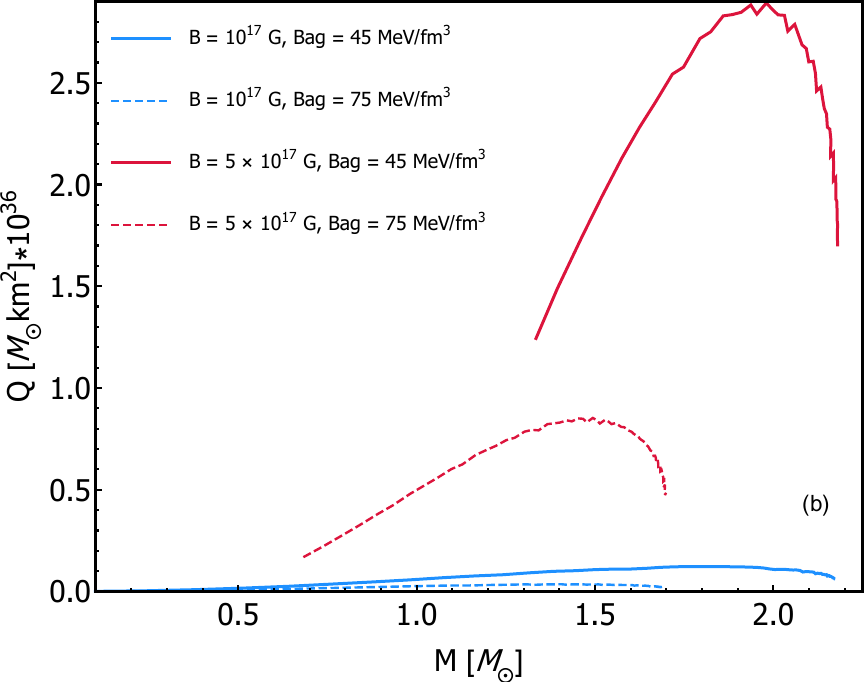}
	\end{tabular}
	\caption{Mass quadrupole $Q$ as function of the mass $M$ at B $=10^{17}$~G and B $=5\times10^{17}$~G for $\mathtt{B_{bag}}=45$~MeV fm$^{-3}$, $\mathtt{B_{bag}}=75$~MeV fm$^{-3}$ \cite{prc}.}
	\label{mqplot}
\end{figure}

The behavior of the mass quadrupole moment as a function of the mass of the stars is shown in Fig.~\ref{mqplot}. The oscillations in the curves are an effect of the presence of Landau levels in the EoS. As can be noticed, $Q$ decreases with $\mathtt{B_{bag}}$, and its maximum values are reached for stars in intermediate mass and deformation region. This behavior is due to the simultaneous dependence of $Q$ on $M$ and $\gamma$, which in particular is determined by the fact that $\gamma$ depends on the EoS and therefore varies among stars. This result differs from the one obtained in Ref. \cite{prc52}, where structure equations derived from the $\gamma$-metric were solved using $\gamma$ as a free parameter. In that case, the highest values of the quadrupole are reached for the most massive stars. Therefore, connecting $\gamma$ with its physical origin directly impacts the observables, and can be used to discriminate between models.

\subsection{Tidal deformability of isolate magnetized Strange Stars}

Pulsars are born from Supernova explosions and are considered rapidly rotating highly magnetized NSs. The most accepted model to explain their very precise rotation period is the lighthouse model which states that the star's rotation and magnetic field axes are not aligned, thus they precess \cite{bonazzola}. Such precession can produce continuous periodical electromagnetic pulses but also gravitational waves (GWs).

We propose a model of static magnetized SSs described by spheroidal structure equations which allow us to connect the deformation with its physical origin as we described in section  \ref{sec2}.  From this model, it is possible to have a mass quadrupole moment to describe the deformation of the star due to the magnetic field. 

But to emit gravitational waves the quadrupole moment has to vary in time. Hence, we adapt the model of Bonazzola \cite{bonazzola} to assume that the isolated magnetized strange star is a slowly rotating object with a quadrupole mass depending on the time. This approximation is suitable to obtain the GW amplitude even for recycled millisecond pulsars because the angular velocity is lower than one,  M$\Omega<0.01$ \cite{ilq81}.

\begin{figure}[h!]
	\centering
	\begin{tabular}{cc}
        \includegraphics[width=0.5\textwidth]{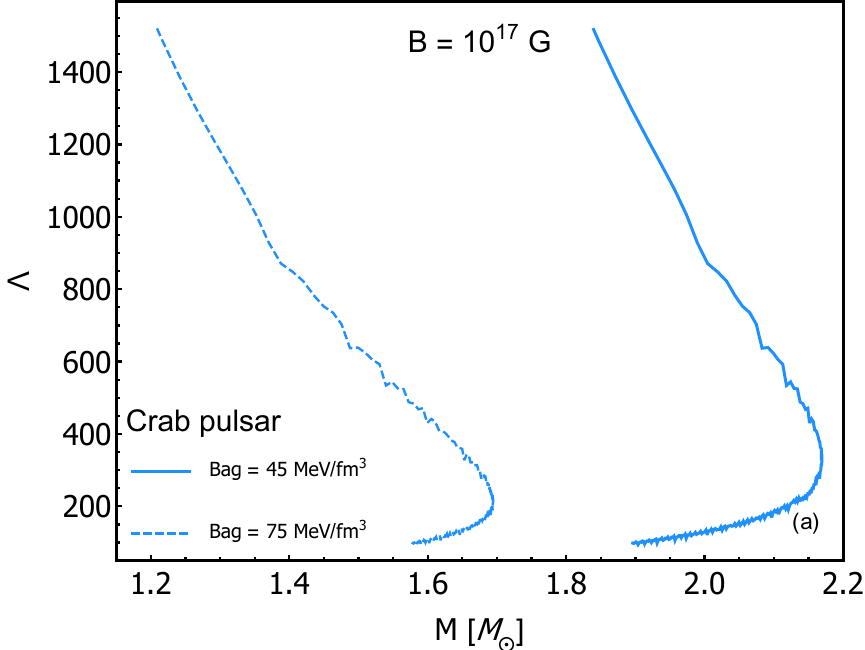}
		& \includegraphics[width=0.5\textwidth]{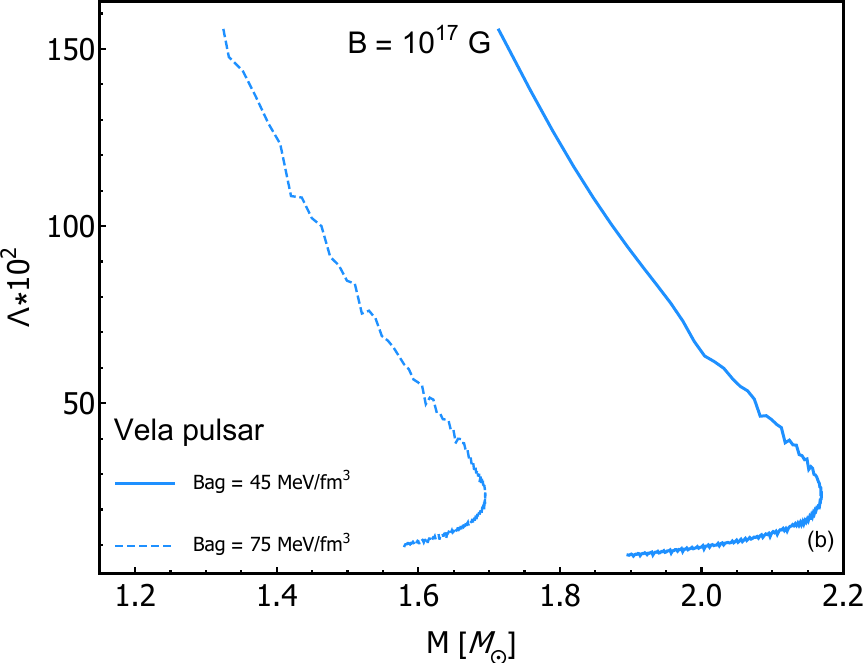}\\
        \includegraphics[width=0.5\textwidth]{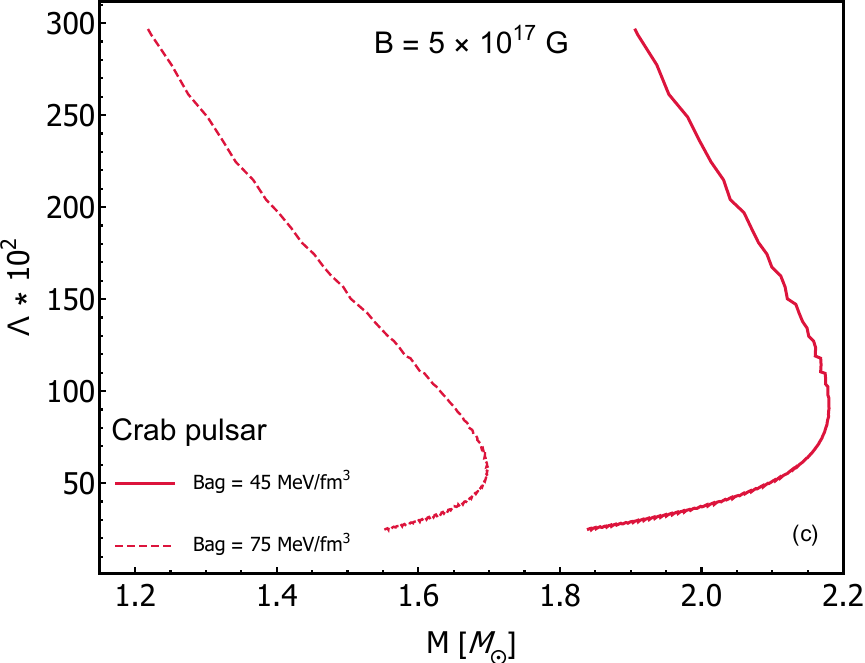}
		& \includegraphics[width=0.5\textwidth]{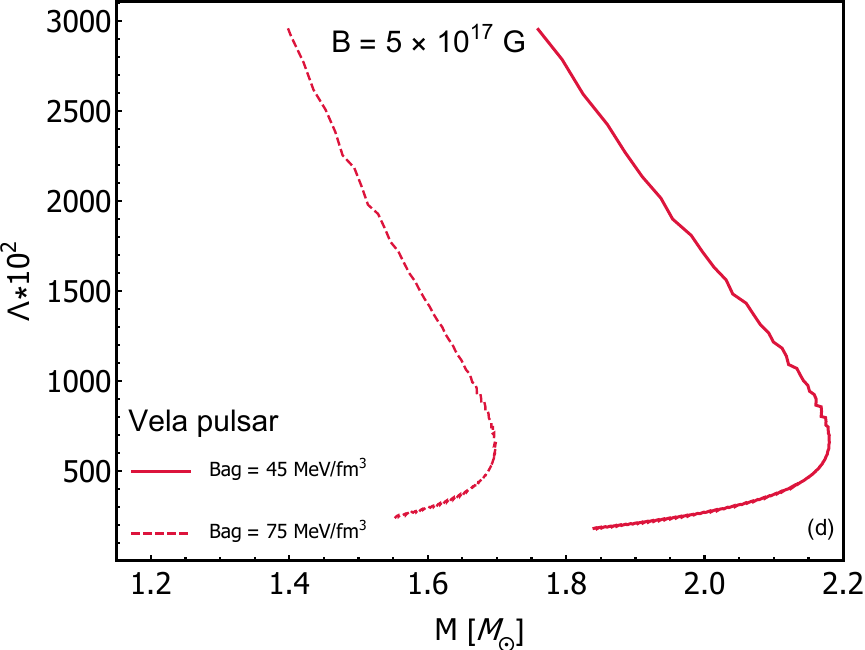}
  \end{tabular}
	\caption{Tidal Deformability for the Crab and Vela pulsars as a function of the mass $M$ at B~$=[10^{17}, 5\times10^{17}]$~G for $\mathtt{B_{bag}}=[45,75]$~MeV/fm$^3$ \cite{prc}. }
	\label{lncm1}
\end{figure}

The kinetic energy of the isolated star is given  $E^{(rot)} =I\Omega^2$ and the variation of this energy is transformed into electromagnetic and gravitational radiation.
Let us consider the quadrupole tidal moment in its Newtonian form $\mathcal{E}^{(B)}=\Omega^2$ \cite{ilq81, ilq86}. The ratio between quadrupole mass and quadrupole \textquotedblleft tidal\textquotedblright\ moment yields  
\begin{equation}
    \lambda^{(B)} = -\frac{Q^{(\gamma)}}{\mathcal{E}^{(B)}}=-\frac{Q^{(B)}}{\Omega^2},
\label{tidalisolate}
\end{equation}
where $\lambda^{(B)}$ is the tidal deformation number for the isolated magnetized star. We calculate this magnitude for the case of the pulsars Crab and Vela, taking into account Eq.~\eqref{mqf} for the mass quadrupole moment. The results are shown in Fig.~\ref{lncm1} where we can see that for a fixed value of the mass, higher values of the tidal deformation are obtained when the magnetic field is increased being the highest of all the ones for Vela.

 \subsection{Gravitational wave amplitude}

 The amplitude of the GW is related to the variation of the properties of the mass of the emitting body employing the following approximate equation, called the quadrupole formula, valid for systems in which the velocities are non-relativistic \cite{h0cita} 
 \begin{equation}
     h_0=\frac{2G}{c^4}\frac{1}{r}\frac{\partial^2Q}{\partial t^2},\label{GWAmp}
 \end{equation}
 where $G$ is the universal gravitational constant, $c$ is the speed of light in a vacuum, $r$ is the distance from the source to the detector, and $Q$ is the quadrupole moment of the source mass distribution time depending, in such way that  $\ddot{Q}=Q^{\gamma}\Omega^2$ is the quadrupole moment of the source mass distribution.

 The term $2G/c^4\sim 10^{-44}$ s$^2$ kg$^{-1}$ m$^{-1}$ in SI units is responsible for the extremely low numerical value of the GW amplitude. From the data collected during the third observation cycle O3 of the Ligo-Virgo Collaboration (LVC) detectors, the lower limit for the amplitude was $\sim 7.6 \times 10^{-26}$, improving by a factor of $1.9$ concerning the previous cycle O2 \cite{Abbott2022}. The LVC started the O4 Observing run on 24 May 2023 \cite{ligo4, bonazzola} aiming for a sensitivity target ranging between 160 and 190 megaparsecs (Mpc) concerning the merging of binary neutron stars \cite{ligo4}.

Following \cite{bonazzola} the  amplitude of the deformation of slow-rotating magnetized quark star yields
\begin{equation}\label{hb}
    h^{Q^{\gamma}}_0=\frac{24\pi^2G}{c^4}\frac{Q^{\gamma}}{P^2r}=
    348\gamma(1-\gamma^2)\left[\frac{M}{M_{\odot}\mathrm{kg}}\right]^3\left[\frac{\mathrm{kpc}}{r}\right]\left[\frac{\mathrm{ms}}{P}\right]^2.
\end{equation}

Note, that in our model the dependence on the magnetic field of the GW amplitude $h_0$, comes from the EoS through the quadrupole moment computed for spheroidal magnetized stars $Q^{\gamma}$. On the contrary, the dependency on the magnetic field of the GW amplitude in \cite{bonazzola} model comes from its geometry, i. e. assuming a dipolar magnetic field by a distortion factor parameter $\beta$. 

\begin{figure}[h]
	\centering
	\begin{tabular}{cc}
		\includegraphics[width=0.5\textwidth]{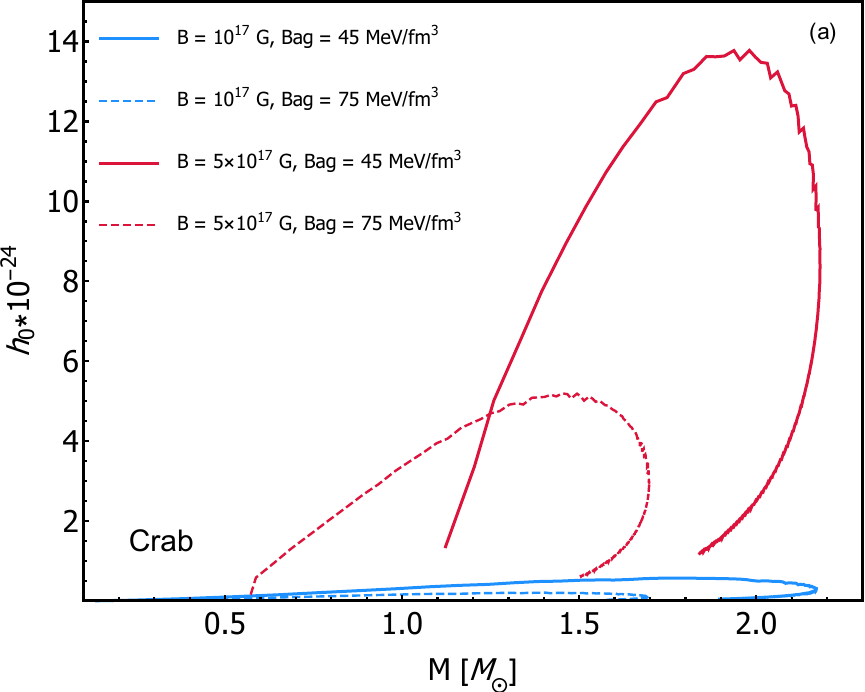}
        & \includegraphics[width=0.5\textwidth]{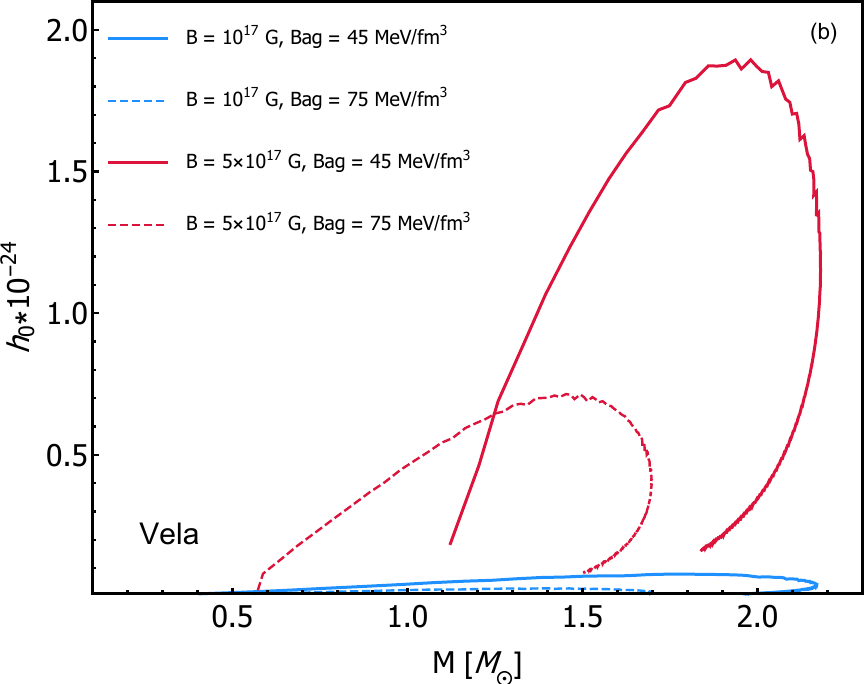}\\
        \includegraphics[width=0.5\textwidth]{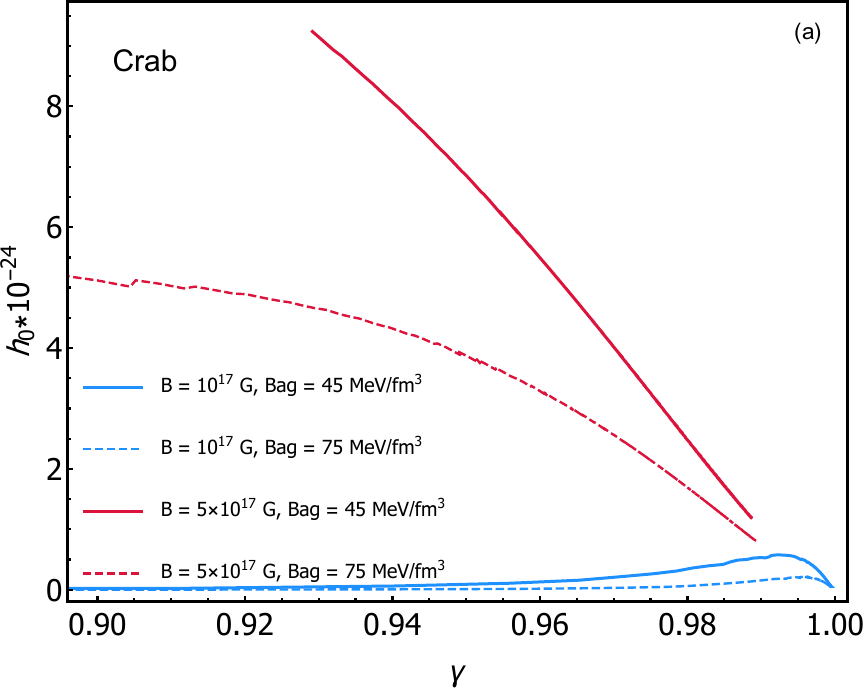}
        & \includegraphics[width=0.5\textwidth]{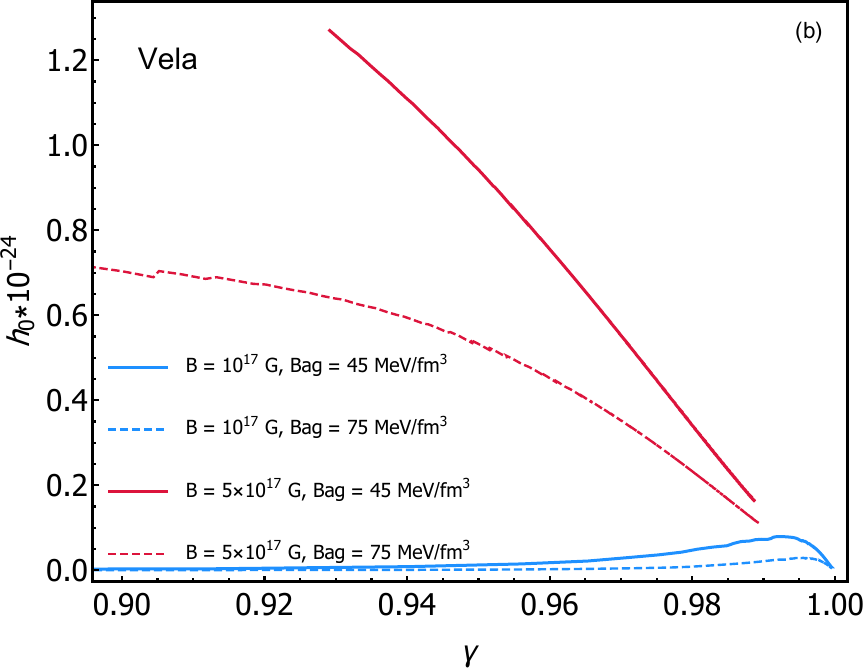}
    \end{tabular}
	\caption{Amplitude of GW as a function of mass (upper panel) and $\gamma$ (lower panel) for Crab and Vela pulsars.}
	\label{lncm}
\end{figure}

To illustrate our model we use the period of Crab and Vela pulsars, which according to \cite{bonazzola} reached the highest values for $h_0$ ($ P = 33$~ms, $\dot{P} = 4.21\times 10^{-13}$, $r=2$~kpc)\cite{Taylor93, Taylor95}. The results can be found in Fig. \ref{lncm} where we have plotted the amplitude as a function of the mass (upper panel) and gamma (lower panel) for B~$=[10^{17}, 5\times10^{17}]$~G. We can see that the amplitude is dictated by the mass quadrupole and is higher for magnetized stars with intermediate mass. Both pulsars show similar curves as a function of mass and gamma. The amplitude corresponding to the Crab pulsar GW is one order higher than the one from the Vela pulsar, due to its rotation period being higher.  
 
\section{Binary system of magnetized Strange Stars}\label{sec3}

Now we study the tidal deformability and Love number for a binary system of two magnetized strange stars.
Tidal fields are originated by the variations of the gravitational field among stellar objects. As a result, deformations appear in the objects due to gravitational interactions.
The tidal field produces many effects. A very well known is caused by the variation of gravity between Earth-Moon-Sun and the tides on the oceans of the Earth.
The relation between tidal field $\mathcal{E}_{ij}$ and the quadrupole moment is given by the equation
\begin{equation}\label{deformacion}
Q_{ij} = \lambda\,\mathcal{E}_{ij}, 
\end{equation}
where $\lambda$ is the tidal deformability and the  Eq (\ref{deformacion})
is the same for a Newtonian or General Relativity description. The tidal deformability can be written in terms of the Love number $k_2$ \cite {Thorne1967, hinderer, Regge1957}.

The relativistic Love number obtained for a spherical binary system  has the form \begin{equation}\label{k2m}
	\begin{split}
		k_2&=\frac{8C^5}{5}(1-2C)^2[2+2C(y_R-1)-y_R]\\
		& \times \left\lbrace \right.   2C(6-3y_R+3C(5y_R-8))+4C^3\left[ \right.   13-11y_R\\
		&+C(3y_R-2)+ 2C^2(1+y_R) \left. \right] \\
		&+3(1-2C)^2[2-y_R+2C(y_R-1)]\ln(1-2C) \left. \right\rbrace ^{-1},
	\end{split}
\end{equation}
where $C=\frac{M}{R}$ is the compactness of the star and  $y= \frac{RH'(R)}{H(R)}$ is obtained from integrating the next equation in the region $0<r<R$
\begin{equation}
	r\frac{dy(r)}{dr}+y(r)^2+y(r)F(r)+r^2Q(r)=0,
\end{equation}
with
\begin{eqnarray}\label{love}
	F(r) &=& \frac{r-4\pi r^3\left[E(r)-P(r)\right]}{r-2M(r)},\\\nonumber
	Q(r) &=& \frac{4\pi r \left(5E(r)+9P(r)+\frac{E(r)+P(r)}{\partial P(r)/\partial E(r)}-\frac{6}{4\pi r^2}\right)}{r-2M(r)}\\\nonumber
	&-&4\left[\frac{M(r)+4\pi r^3P(r)}{r^2(1-2M(r))/r}\right]^2.
\end{eqnarray}

As the tidal deformability depends directly on the mass and radius of the star, it could be useful to verify or discard models of neutron stars.

On the other hand the signal of the event GW170817 has bounded the tidal deformability so, we can take for compactness our MR solutions of spherical structure equations and we bound the tidal deformability with the constrained by GW170817 event. This procedure allows us to study the reliability of our model, in particular the range of validity of microscopic parameters of the EoS.

To do that we solve the spheroidal structure equation together with the Love number Eq.~(\ref{love}).
The latter corresponds to those assuming spherical symmetry and not considering spheroidal symmetry. However, this approximation is reasonable since our model is only valid for small deviations of the spherical shape, i.e., $\gamma\sim 1$,

Once $k_2$ is computed, the dimensionless tidal deformability is given by
\begin{equation}\label{Lambda}
\Lambda = \frac{2}{3}k_2\frac{R^5}{M^5}=\frac{2}{3}k_2C^{-5}.
\end{equation}

This study represents the first step in considering the effects of the magnetic field on tidal deformations. For the bigger picture, we aim to obtain the Love number from the $\gamma$ metric, perturbing it due to the magnetic field \cite{rezzolla}.

\begin{figure}[h]
	\centering
	\begin{tabular}{cc}
		\includegraphics[width=0.5\textwidth]{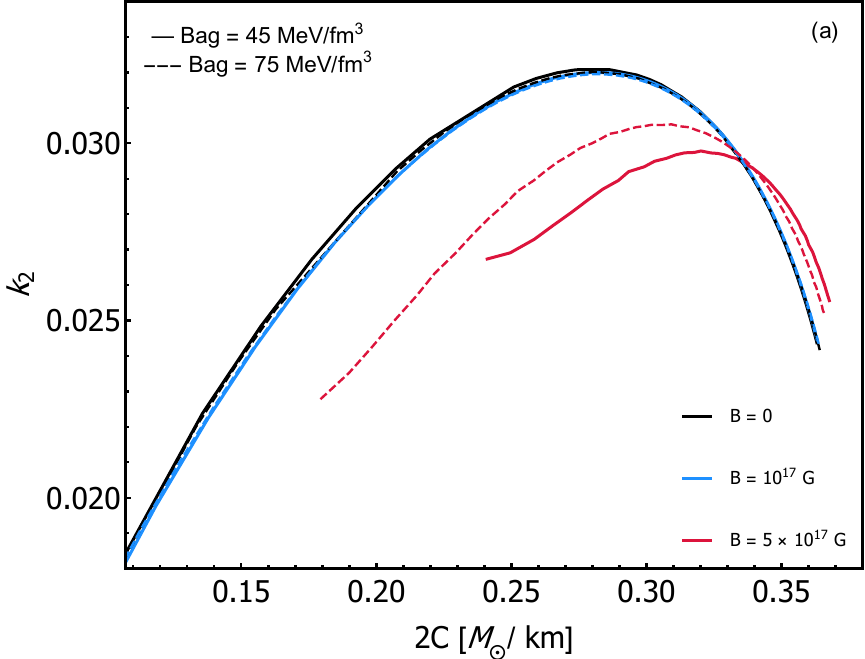}
		& \includegraphics[width=0.5\textwidth]{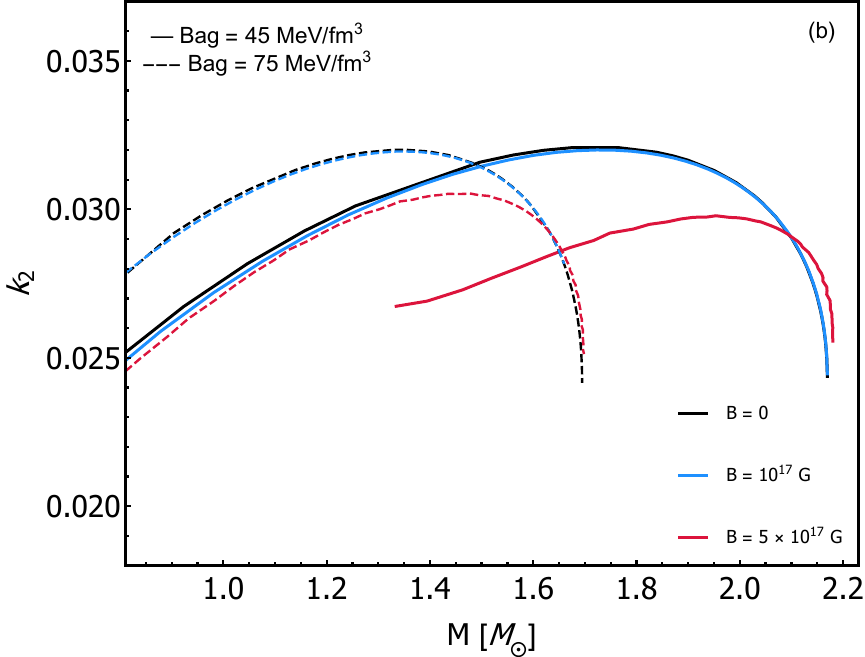}
	\end{tabular}
	\caption{ Love number  $k_2$ as a function of compacity $C$ (a) and mass $M$ (b) for $B=0$, $B=10^{17}$~G and $B=5\times10^{17}$~G.}
	\label{lncm2}
\end{figure}

 In Fig.~\ref{lncm2} we plot the Love number based on compactness and star mass. The Love number gives a measure of how easily a star can be deformed. If most of the mass of the star is concentrated in the center, the mass deformation will be smaller. For the same value of the magnetic field, the Love number increases with the $\mathtt{B_{bag}}$, while for the same value of the $\mathtt{B_{bag}}$ decreases with the magnetic field. In addition, the Love number decreases by increasing compactness. The effects of the magnetic field are remarkable, because for $B<5\times 10^{17}$~G the behavior is practically equal to the case of the non-magnetic field model. By comparing our results with other models of stars with quark matter \cite{Flores}, we observed that the Love number is approximately in the same range as many of the results obtained although it reaches slightly smaller values.
\begin{figure}[h]
	\centering
	\begin{tabular}{cc}
		\includegraphics[width=0.5\textwidth]{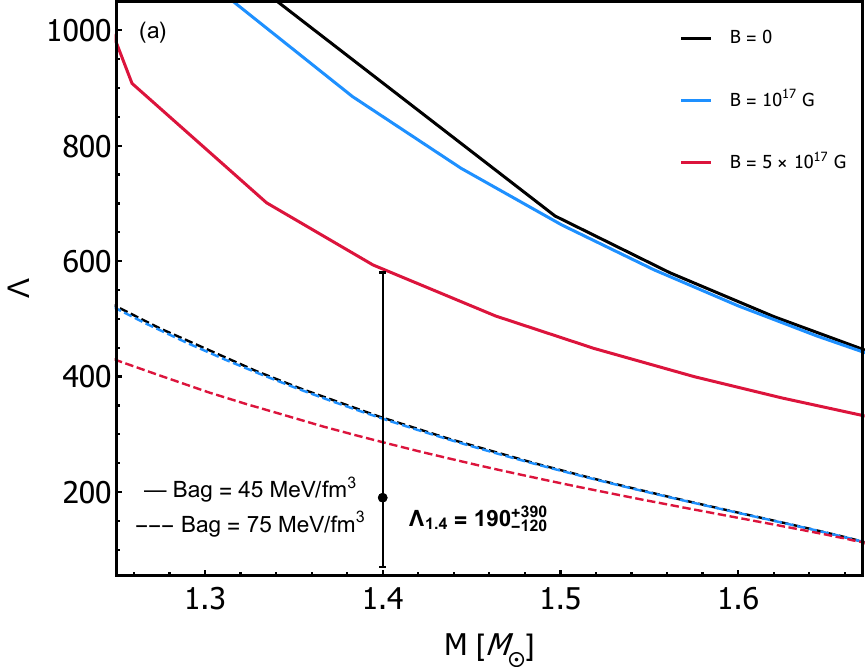}
		& \includegraphics[width=0.5\textwidth]{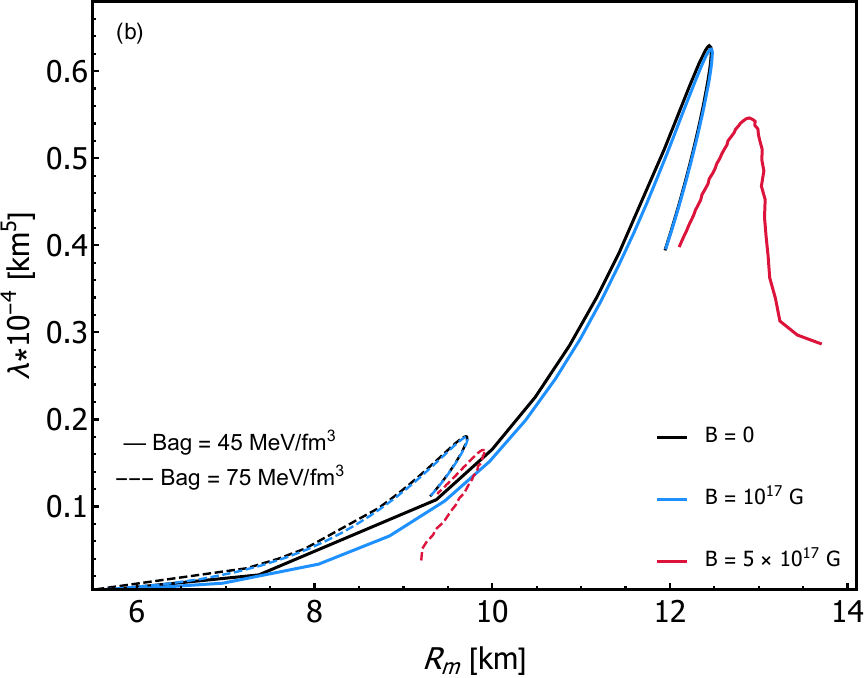}
	\end{tabular}
	\caption{Dimensionless tidal deformation  $\Lambda$ as a function of mass $M$ (a) and tidal deformation $\lambda$~[km]$^5$ as a function of mean radius  $R_m$ (b) at $B=0$, $B=10^{17}$~G and $B=5\times10^{17}$~G compared to the result  $\Lambda_{1,4}=190^{+390}_{-120}$ obtained by the Ligo-Virgo Collaboration \cite{asna22, Abbott2017f}.}
\label{Lm}
\end{figure}

In Fig.~\ref{Lm} (a) the dimensionless tidal deformation for a magnetized binary system is represented according to its mass. For the same value of $\mathtt{B_{bag}}$, increasing the magnetic field decreases $\Lambda$, while this increases for smaller values of $\mathtt{B}_{bag}$. We must highlight that the curves for $\mathtt{B_{bag}}=75~MeV/fm^3$ are within the set range for the observational parameter obtained from the GW170817, confirming that our results are consistent with the observations and also with other quarks matter works in the literature. In addition, in Fig. \ref{Lm} (b) we show how $\lambda$ depends on the mean radius. We can verify that the same characteristics observed in Fig. \ref{Lm} (a) are presented in the curves $\lambda$ vs $R_m$, i.e., that $\lambda$ increases by decreasing the $\mathtt{B_{bag}}$ and the magnetic field.
\begin{figure}[h!]
	\centering
	\begin{tabular}{c}
		\includegraphics[width=0.5\textwidth]{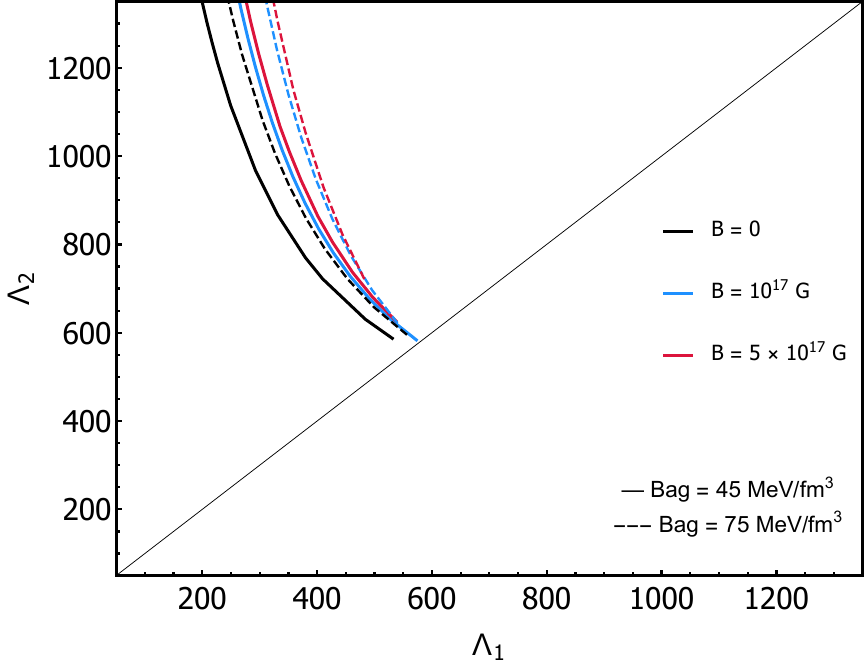}
	\end{tabular}
	\caption{Dimensionalness tidal deformability for the magnetized binary system of strange stars constrained by GW170817 .}
	\label{l12}
\end{figure}

We want to compute also the $\Lambda_2$. To do that we consider that in the final inspire phase of a binary system, the  GW is characterized by a  $\tilde{\Lambda}$ that is expressed in function of the two masses of the binary system as
\begin{equation}\label{lm}
    \tilde{\Lambda} = \frac{16}{13}\frac{(M_1+M_2)M_1^4\Lambda_1+(M_2+12M_1)M_2^4\Lambda_2}{(M_1+M_2)^5},
\end{equation}
where $\Lambda_1$ and $\Lambda_2$ are the dimensionaless tidal deformations of each star defined in \eqref{Lambda}. This result was obtained for the first time in \cite{hinderer} and is used to investigate the response of stellar material to the tidal field by extracting directly from the observed wavelength. The mass of the second star, $M_2$, is related to $M_1$ through the resulting mass defined as \cite{Abbott2017f}
\begin{equation}\label{mm}
\mathcal{M} = \frac{\left(M_1M_2\right)^{3/5}}{\left(M_1+M_2\right)^{1/5}}.
\end{equation}
$\mathcal{M}$ is the chirp mass and we also consider it as the obtained from the GW170817 event.
The predicted values for masses given by\cite{Abbott2017f} are $\mathcal{M}=1.188^{+0.004}_{-0.002}$~$M_{\odot}$ \cite{Abbott2017f}, generating a variation of $1.17\leq M_2/M_{\odot}\leq 1.36$\cite{Abbott2017f, Abbott2018c} for the mass of the companion star. The obtained data allowed LVC to set some restrictions on $\Lambda_1$ and $\Lambda_2$, in addition to determining a range for $ \Lambda_{1,4}$ (a star's tidal deformation with $M=1,4$~M$_{\odot}$).

The $\Lambda_1$ and $\Lambda_2$ tidal deformations of the magnetized SSs binary system are displayed in Fig.~\ref{l12}, where $\Lambda _2$ was calculated from Eq.~\eqref{lm}, replacing the previously calculated values for $M_1 $ and $\Lambda_{1}$. The behavior of the curves is compatible with other matter works of quarks, obtaining slightly smaller values for $\Lambda_2$.

\section{Conclusions}

We have studied the magnetic field effects on the tidal deformation of isolated magnetized SSs and the GW amplitude generated for this star. On the other hand, we have studied the tidal deformation associated with the inspiral phase of a binary magnetized SSs imposing constraints from the data by the event GW170817.

For the isolated slow-rotating stars, higher values of tidal deformation are obtained for a fixed
value of the mass when the magnetic
field is increased being the highest of all the ones for Vela.
The amplitudes of GW of magnetized isolated SSs are of the order of $10^{-24}$, $10^{2}$ smaller than the produced by the crash of the binary system.  The amplitude is three orders higher than those estimated for the Crab pulsar by Bonnaloza et al with a different dependency on the magnetic field as a consequence of the magnetized star being described by anisotropic EoS and consequently spheroidal structure equation.
This highlights the importance of taking models for the structure equations that include the effects of the magnetic field.

The Love number and the tidal deformation were calculated for one component of the binary system with mass $M_1$, obtained from the solutions of Eqs. $\gamma$. Then, from the inferred data from the GW170817 event for the chirp mass of the collision and tidal deformation of the star,  the mass and tidal deformation of the second component of the binary system were calculated. 

The Love number $k_2$ of the component of the binary system with mass $M_1$ decreases with the magnetic field and increases with the $\mathtt{B_{Bag}}$, while it decreases with increasing compactness. The magnetic field notably modifies the behavior of $k_2$ for values greater than $5\times10^{17}$~G. When comparing our results with other studies for EoE with matter of quarks reported in the literature, we conclude that they are within the range of estimated values for $k_2$. The tidal deformation $\Lambda_1$ of the component of the binary system with mass $M_1$ decreases with the increase in the magnetic field, while it is greater for smaller values of the parameter $\mathtt{B_{Bag}}$ of the EoE for a fixed value of the magnetic field. In particular, the curves for $B_{\mathtt{Bag}}=75 MeV/fm^3$ are within the observational range inferred from event GW170817. Note, that the isolated stars show a different behavior. Increasing the magnetic field or the Bag parameter of the model implies obtaining higher values of deformability of the star $\Lambda$.
From the expressions for the mass and the deformation of the star resulting from the collision and the data inferred from the GW170817 event for these, the tidal deformation $\Lambda_2$ of the second component star of the binary system was calculated. The values obtained for $\Lambda_2$ agree with those reported in the literature by other studies for EoE with matter of quarks. In particular, our results agree with the model proposed in \cite{vivian} for non-magnetized EEs that consider, in addition to foreign matter, the presence of gluons.

\section*{Acknowledgments}

The authors thank Adriel Rodríguez and Duvier Suarez Fontanella for the fruitful comments and discussions. The authors have been supported by  PNCB-MES Cuba No. 500.03401 and by the grant of the ICTP Office of External Activities through NT-09. 

\bibliographystyle{unsrt}  
\bibliography{MagnetizedStrangeStarsGWs}

\end{document}